\documentclass[11pt]{amsart}

\usepackage{ulem} 
\usepackage[english]{babel}
\usepackage[latin1]{inputenc}
\usepackage{bm}
\usepackage{fullpage}
\usepackage{amssymb}
\usepackage{amsmath}
\usepackage{latexsym}
\usepackage{graphicx}

\newtheorem{theorem}{Theorem}
\newtheorem{proposition}{Proposition}
\newtheorem{lemma}{Lemma}

\newtheorem{assumption}{Assumption}
\newtheorem{claim}{Claim}

\newcommand*{\ligne}[1][1]{\rule[0.4em]{0.5\textwidth}{0.5pt}\par}

\author{Frédéric Meunier}
\address{Université Paris Est, LVMT, ENPC, 6-8 avenue Blaise Pascal, Cité Descartes
Champs-sur-Marne, 77455 Marne-la-Vallée cedex 2, France.}
\email{frederic.meunier@enpc.fr}
\author{Nicolas Wagner}
\address{Université Paris Est, LVMT, ENPC, 6-8 avenue Blaise Pascal, Cité Descartes
Champs-sur-Marne, 77455 Marne-la-Vallée cedex 2, France.}
\email{nicolas.wagner@gmail.com}

\title{Dynamic assignment: there is an equilibrium !}

\begin{document}

\begin{abstract}
Given a network with a continuum of users at some origins, suppose that the users wish to reach specific destinations, but that they are not indifferent to the time needed to reach their destination. They may have several possibilities (of routes or deparure time), but their choices modify the travel times on the network. Hence, each user faces the following problem: given a pattern of travel times for the different possible routes that reach the destination, find a shortest path.

The situation in a context of perfect information is a so-called Nash equilibrium, and the question whether there is such an equilibrium and of finding it if it exists is the so-called equilibrium assignment problem. It arises for various kind of networks, such as computers, communication or transportation network.

When each user occupies permanently the whole route from the origin to its destination, we call it the static assignment problem, which has been extensively studied with pioneers works by Wardrop or Beckmann \cite{Wa52,Be56}. A less studied, but more realistic, and maybe more difficult, problem is when the time needed to reach an arc is taken into account. We speak then of a dynamic assignment problem. 
 
Several models have been proposed. For some of them, the existence of an equilibrium has been proved, but always under some technical assumptions or in a very special case (a network with one arc for the case when the users may chose their departure time). 

The present paper proposes a compact model, with minimal and natural assumptions. For this model, we prove that there is always an equilibrium. To our knowledge, this imply all previous results about existence of an equilibrium for the dynamic assignment problem.
\end{abstract}

\maketitle

\section{Introduction}

\subsection{Dynamic equilibrium assignment}  Consider over a time interval $I:=[0,H]$, say a day, a network where a set of users travel along
directed paths, called routes, connecting origins to destinations. We denote by $R$ the set of routes. At the beginning of the day users are at origins and wish to reach a specific destination by the end of the day.
In order to do so, they take a travel decision on the network, i.e. choose a route and a departure time on this route. Yet users decisions depends on route travel time over the network, itself depending on the flow of users following each routes and thus on the decisions of the other users.

Finding an equilibrium (in the Nash sense) of such a problem is, roughly speaking, the {\em dynamic equilibrium assignment} problem. Contrary to the {\em static assignment} (studied by Wardrop \cite{Wa52}, Beckmann \cite{Be56} Patricksson \cite{Pa94}, Roughgarden and Tardos \cite{RoTa02}, Milchtaich \cite{Mi05} and many others), where users occupy the whole route permanently, the {\em dynamic assignment} problem represents explicitly the time needed to reach any arc in the network, thus giving a more realistic description of the traffic propagation on the network. Different models have been proposed for the dynamic assignment problem (Vickrey \cite{VIC69}, Merchant and Nemhauser \cite{MeNe78}, Friesz and al. \cite{FBS93} Lindsey \cite{LIN04}, ...) -- all in the context of transport, which contrasts with the static assignment which is well studied for communication and computers networks as well. The present paper presents a general framework for the study of Nash equilibrium over a network and show that it can easily be used to prove the existence of a dynamic assignment.

In the paper, each flow of users entering a route or an arc is modeled as a measure. For a subset $J$ of instants, the measure of $J$ is the number of users whose entrance time into the route or the arc is in $J$. Each flow of user entering a route is a measure on $I$  but each flow of users entering an arc is modeled as a measure on $\mathbb{R}$. Indeed, for the arcs, we have less control on the entrance time.
These sets of measures will be denoted respectively by $\mathcal{M}(I)$ and $\mathcal{M}(\mathbb{R})$. The set of continuous maps from $\mathbb{R}$ to itself (for instance: entrance time $\rightarrow$ exit time, which may exceed the bounds of $I$) is denoted by $\mathcal{C}(\mathbb{R})$. 

\subsubsection{Supply} We have a network modeled as a directed graph $G=(V,A)$. Each arc is endowed with an {\em arc travel time function} $t_a$ that tells the time needed to traverse the arc $a$ for each entrance time $h$, when we know the flow $X$ that enters $a$. For a particular $h$ and a particular flow $X$, we denote this time by $t_a(X)(h)$. Hence, $t_a$ is a $\mathcal{M}(\mathbb{R})\rightarrow\mathcal{C}(\mathbb{R})$ map. To be consistent with the nature of a transportation network, we will make five very natural assumptions on $t_a$ denoted continuity, causality, strict fifoness, no infinite speed and finiteness (see Subsection \ref{subsec:time}). Suppose that all the users have made a choice. This collection of choices is modeled by a measure $\boldsymbol{X}=(X_r)_{r\in R}$ on $R\times I$, which encodes for each pair $(r,h)$ the flow of users entering the route $r$ at time $h$. These flows induces {\em route travel time functions} $(t_r(\boldsymbol{X}))_{r\in R}$ as follows :

Let $r=a_1a_2\ldots a_n$ be a route, $t_r(\boldsymbol{X})(h)$, which is the time needed to travel through the whole route $r$ when entering the route at $h$, is the sum of the times needed to traverse the arcs $a_1,a_2,\ldots,a_n$: 
\begin{equation}
\label{eq:tr}
t_r(\boldsymbol{X})(h):=\sum_{i=1}^nt_{a_i}(Y_{a_i})(h_i)\quad\mbox{with }h_1:=h\mbox{ and }h_{i+1}:=h_i+t_{a_i}(Y_{a_i})(h_i)\mbox{ for }i=1,\ldots,n-1,
\end{equation}
 where $Y_{a_i}:=\phi_{a_i}(\boldsymbol{X})$ is the flow entering $a_i$ (we call these flows the {\em outflow} of $\boldsymbol{X}$) and that results from all the $X_r$ such that route $r$ contains $a_i$. Note that the way to deduce the flows $Y_{a_i}$ on the arcs from the flows $X_r$ on the routes -- encoded by the map $\phi_{a_i}$ -- is not straightforward and need further precisions that we will be given in Subsection \ref{subsec:flowm}. Let us already say that they are consequences of the nature of the arc travel time functions $t_a$. 
Denoting by $\mathcal{T}_r$ the set of maps from the set of measures on $R\times I$ to the set of continuous maps $\mathcal{C}(\mathbb{R})$, we see that $t_r$ is an element of $\mathcal{T}_r$. Finally, let us denote $\mathcal{T}:=\prod_{r\in R}\mathcal{T}_r$. An element $\boldsymbol{t}=(t_r)_{r\in R}$ of $\mathcal{T}$ is a so-called {\em route travel times pattern}.

\subsubsection{Demand} The users make their choices according to their utility functions $u: R\times I\rightarrow\bar{\mathbb{R}}$ where $\bar{\mathbb{R}}$ is the set of extended reals. The utility functions in the paper are supposed to be upper semicontinuous (see the definition in Section \ref{sec:tools}). We denote the space of these utility functions by $\mathcal{S}_{R\times I}$.

Then, each user is identified by a continuous function $\hat{u}$ from the set of possible route travel times patterns $\mathcal{T}$ to $\mathcal{S}_{R\times I}$: given $\boldsymbol{t}=(t_r)_{r\in R}$ a route travel times pattern, $\hat{u}(\boldsymbol{t})$ is a (upper semicontinuous) utility function that will guide the user in his choice. Given a route $r$ and a departure time $h$ chosen by the user and given a route travel time pattern $\boldsymbol{t}$, which depends on the other users, $\hat{u}(\boldsymbol{t})(r,h)$ represents the utility enjoyed by the user characterized by $\hat{u}$. Working with semicontinuous functions allows to cover many distinct situations. For instance, as presented in Section \ref{sec:applications}, we can force users to depart at given time, hence restricting their travel choice to a route choice. 


We denote by $\mathcal{C}(\mathcal{S}_{R\times I})$ the space of such functions $\hat{u}$, endowed with the compact-open topology (see Section \ref{sec:tools}). The distribution of users is modeled by a (Borel) measure $U$ on $\mathcal{C}(\mathcal{S}_{R\times I})$. The {\em total number of users} is defined by $N:=U(\mathcal{C}(\mathcal{S}_{R\times I}))$.

\subsubsection{Equilibrium} Finally, we need to formulate a {\em Nash equilibrium} in this context. We follow Khan (\cite{KHA89}).
Let $M$ be a measure a product $E\times F$. We use the notation $M_E$ to denote the {\em marginal} of $M$ on $E$, that is, for $E'\subseteq E$, measurable, we have $M_E(E')=M(E'\times F)$.
\medskip

\fbox{
\begin{minipage}{0.9\textwidth}
\textbf{Nash Equilibrium}

\ligne
A (Borel) measure $D$ on $R\times I\times\mathcal{C}(\mathcal{S}_{R\times I})$ is a {\em Nash equilibrium} if  
\begin{enumerate}
\item $D_{\mathcal{C}(\mathcal{S}_{R\times I})}=U$.
\item $D\big\{(r,h,\hat{u})\in R\times I\times\mathcal{C}(\mathcal{S}_{R\times I}): 
\,\hat{u}(\boldsymbol{t})(r,h)\geq\hat{u}(\boldsymbol{t})(r',h')\mbox{ for all }r'\in R, h'\in I\big\}=N$ with $\boldsymbol{t}:=\boldsymbol{t}(D_{R\times I})$.
\end{enumerate}
$D_{R\times I}$ is interpreted as a vector $\boldsymbol{X}=(X_r)_{r\in R}$ of users flows entering the routes in $R$, by the identification $X_r:=D_{R\times I}(\{r\}\times I)$. 
\end{minipage}}
\medskip

Essentially, the formulation of the equilibrium states that the volume of users with an optimal travel decision regarding the route travel times pattern $\boldsymbol{t}(\boldsymbol{X})$ is the total volume of users. 

\subsection{Main results} The purpose of the present paper is twofold. First, we present a very general equilibrium model with very natural assumptions. Second, the existence theorem stated below is established. In its proof, we use a theorem by Khan, rephrased below in Subsection \ref{subsec:kh}. Our main technical contribution consists in proving that whenever the arc travel time functions $t_a$ satisfy the natural assumptions of Subsection \ref{subsec:time}, the route travel time functions $t_r$ are continuous from $\mathcal{M}(I)$ to the set $\mathcal{C}(I)$, the crucial fact that allows to use Khan's theorem.

\begin{theorem}\label{thm}
Given a directed graph $G=(V,A)$ with arc travel time functions $(t_a)_{a\in A}$ satisfying Assumptions \ref{ass:cont}-\ref{ass:causality} and given a measure $U$ on the set of possible users (identified with $\mathcal{C}(\mathcal{S}_{R\times I})$), there is a Nash equilibrium.
\end{theorem} 

This theorem solves an open question in transportation science, since it has been asked for many years whether there exists an equilibrium for a general model of dynamic traffic assignment (see Friez et al \cite{FBS93}).

Note moreover that our definition of the utility functions $u$ as upper semicontinuous functions, allows a lot of variations: users can or cannot choose their $o$-$d$ pair, they can or cannot choose their departure time, we can arbitrarily put a bonus or a malus on some routes, or on some set of time departures, and so on... .

Finally Theorem \ref{thm} contains previous equilibrium results for dynamic assignment, such as which of Zhu and Marcotte \cite{ZHM00}, or Mounce \cite{MOU06}, or Lindsey \cite{LIN04}.

All these consequences are discussed in Section \ref{sec:applications}.

\subsection{Organization of the paper}

Section \ref{sec:tools} gives the main tools and notations of the paper. Since continuity results will be the main technical aspects of our work, we will carefully define in this section the topologies of our different sets. Moreover, Section \ref{sec:tools} gives a theorem of Khan (Theorem \ref{thm:kh}) that roughly speaking shows the existence of an equilibrium as soon as the functions $t_r$ are continuous. In Section \ref{sec:model}, we define the model: given the flows $X_r$ that enter the routes, how can we build the flow on the arcs ? The result of such a process is the \emph{outflow} of the route flows $\boldsymbol{X}$, and its properties are given by Proposition \ref{prop:routearc}. 
To prove Theorem \ref{thm}, it is sufficient to prove that Assumptions \ref{ass:cont}-\ref{ass:causality} on the arc travel time functions $t_a$ implies the continuity of the $t_r$. This is the purpose of Section \ref{sec:cont}, with the help of Proposition \ref{prop:routearc}. In the last section (Section \ref{sec:applications}), we will see how it covers previous results in the area of dynamic traffic assignment.

\section{Main tools and notations}

\label{sec:tools}



\subsection{Sets and topologies} We use $\mathcal{M}(E)$ to denote the set of measures on a set $E$. We will systematically use the weak convergence topology on any set of measures encountered in the paper. 

\subsubsection{Upper semicontinuity}Assume that $E$ is a nonempty and compact Hausdorff space. A function $u:E\rightarrow\bar{\mathbb{R}}=\mathbb{R}\cup\{-\infty,+\infty\}$ is said to be upper continuous if the {\em hypograph} of $u$ is closed. The hypograph of a function $f:E\rightarrow\bar{\mathbb{R}}$ is the set $\{(x,y)\in E\times\bar{\mathbb{R}}:f(x)\geq y\}$. We denote by $\mathcal{S}_E$ the space of upper semicontinuity function $E\rightarrow\bar{\mathbb{R}}$.

\subsubsection{Hypotopology} The space $\mathcal{S}_E$ is endowed with the {\em hypotopology} where to maps are ``close'' if their hypographs are ``close'' for the Hausdorff distance. 

\subsubsection{Weak convergence of measures}A sequence of measure $M_n$ defined on a set $E$ is said to weakly converge toward a measure $M$ on $E$ if
\begin{itemize}
\item[(i)]$\limsup_{n\rightarrow+\infty}M_n(F)\leq M(F)$ for any closed subset $F$ of $E$, and
\item[(ii)] $\limsup_{n\rightarrow+\infty}M_n(E)=M(E)$.
\end{itemize}

For more informations about the weak convergence, see \cite{Top70}.

\subsubsection{Compact-open topology} Let $E$ and $F$ be two topological spaces, and let $\mathcal{C}(E, F)$ denote the set of all continuous maps between $E$ and $F$. Given a compact subset $K$ of $E$ and an open subset $O$ of $F$, let $V(K, O)$ denote the set of all functions $f$ in $\mathcal{C}(E, F)$ such that $f(K)$ is contained in $O$. Then the collection of all such $V(K, O)$ is a subbase for the {\em compact-open topology}, that is the compact-open topology is the smallest topology containing all such $V(K, O)$.

\subsubsection{Topology for sets of continuous mappings} Denote by $\mathcal{C}(I)$ the set of continuous map from $I$ to $\mathbb{R}$. We endow $\mathcal{C}(I)$ with the supremum norm. 

\subsubsection{Restriction of measures} Let $M$ be a measure on $\mathbb{R}$. For any $h\in\mathbb{R}$, we denote by $M|_h$ the measure  such that $M|_h(J):=M(J\cap]-\infty,h])$ for all measurable subsets $J$ of $\mathbb{R}$. We extend this notation to the measure on $R\times\mathbb{R}$. If $M$ is such a measure, $M|_h(R'\times J)=M(R'\times(J\cap]-\infty,h]))$ for all measurable subsets $J$ of $\mathbb{R}$ and all subsets $R'$ of $R$. 

\begin{claim}\label{cl:rest}
If $h_2>h_1$, then for any measure $M$, we have $M|_{h_1}=M|_{h_2}|_{h_1}$.
\end{claim}

The proof is straightforward.

\subsection{Khan's theorem}\label{subsec:kh}
In 1996, generalizing an approach by Mas-Colell (\cite{MAS84}), Khan has proposed a theorem that gives a sufficient condition for the existence of a Nash equilibrium with a continuum of users \cite{KHA89}. Khan's theorem rephrased in our context is :

\begin{theorem}\label{thm:kh} Let $U$ be a measure on $\mathcal{C}(\mathcal{S}_{R\times I})$.
Suppose that the functions $t_r$ are continuous for every $r\in R$ (as function from $\mathcal{M}(I)$ to $\mathcal{C}(\mathbb{R})$). Then there exists a Nash equilibrium.
\end{theorem}

\section{Model}\label{sec:model}

\subsection{Arc travel time function} 

\label{subsec:time}

We want to apply Theorem \ref{thm:kh} to prove Theorem \ref{thm}. In order to do that, we need to prove that the route travel time functions $t_r:\mathcal{M}(I)\rightarrow\mathcal{C}(\mathbb{R})$ are continuous. The $t_r$ derive from the arc travel time functions $t_a$, defined for each arc $a\in A$ of $G$. To establish the continuity of the functions $t_r$, we need five (very natural) assumptions of the nature of $t_a$. 
\medskip

\fbox{
\begin{minipage}{0.9\textwidth}
\textbf{Assumptions on arc travel time $t_a$}

\ligne 

\begin{assumption}\label{ass:cont}\textup{[Continuity]}
$t_a:\mathcal{M}(\mathbb{R})\rightarrow\mathcal{C}(\mathbb{R})$ is continuous.
\end{assumption}

\begin{assumption}\label{ass:noinfspeed}\textup{[No infinite speed]}
There exists $t_{\min}>0$ such that for all $Y\in\mathcal{M}(\mathbb{R})$ and all $h\in\mathbb{R}$, we have $t_a(Y)(h)>t_{\min}$.
\end{assumption}

\begin{assumption}\label{ass:end}\textup{[Finiteness]} There exists a continuous map $t_{\max}:\mathbb{R}_+\rightarrow\mathbb{R}_+$ such that $t_a(Y)(h)\leq t_{\max}(Y(\mathbb{R}))$ for all $h\in\mathbb{R}$.
\end{assumption}

\begin{assumption}\label{ass:strictfifo} \textup{[Strict Fifoness]} Let $h_1<h_2$ in $\mathbb{R}$ and let $Y\in\mathcal{M}(\mathbb{R})$. Whenever $Y[h_1,h_2]\neq 0$, we have $h_1+t_a(Y)(h_1)<h_2+t_a(Y)(h_2)$.
\end{assumption}

\begin{assumption}\label{ass:causality}\textup{[Causality]}
For all $h\in\mathbb{R}$ and $Y\in\mathcal{M}(\mathbb{R})$, we have $t_a(Y|_h)(h)=t_a(Y)(h)$.
\end{assumption}
\end{minipage}
}
\medskip

Given $t_a$, we introduce the \emph{arc exit time} function:
\begin{equation}
\label{eq:Ha}
H_a(Y)(h):=h+t_a(Y)(h)\quad\mbox{for }Y\in\mathcal{M}(\mathbb{R})\mbox{ and } h\in \mathbb{R}
\end{equation}

$H_a(Y)\in\mathcal{C}(\mathbb{R})$ is the map that takes en entrance time on arc $a$ and returns the exit time $H_a(Y)(h)$, when we know the flow $Y$ that goes through the arc $a$. Remember that we define flows on arcs as measure on $\mathbb{R}$, and not $I$, although the users enter the routes in the time interval $I$, they don't necessarily leave them in $I$. 

Assumption \ref{ass:cont} is a very classical assumption: a small variation of flow leads to a small variation of the arc travel time. Assumption \ref{ass:noinfspeed} can be restated as for all $Y\in\mathcal{M}(\mathbb{R})$ and all $h\in\mathbb{R}$, we have $t_a(Y)(h)>t_{\min}$. It amounts to say that the time needed to go through an arc is bounded from below. It is natural since otherwise it would mean that there are arcs on which users can have infinite speeds. The finiteness condition (Assumption \ref{ass:end}) assumes that if we wait for a sufficient long time, a flow of users on any arc $a$ leaves completely the arc. The fifo condition (\ref{ass:strictfifo}) is a reformulation of a standard assumption in transportation science. Unformally, it means that if two cars enter an arc in a given order, they depart the arc in the same order. Finally Assumption \ref{ass:causality} simply implies that the time needed to go through an arc depends on the traffic before the entrance time, but not on the traffic that follows the entrance time.

\subsection{Flowing function} \label{subsec:flows}

The main object presented in this subsection is the {\em flowing function} that, given the collection of flows entering an arc $a$ (seen as an element of $\mathcal{M}(R \times\mathbb{R})$), returns the collection of flows leaving the arc $a$. This flowing function is denoted by $\psi_a$. Note that it is entirely defined from the arc exit time function $H_a$, itself entirely defined form the arc travel time function, and hence, that the behavior of a the network is entirely contained in the definitions of the graph $G=(V,A)$ and of the arc travel time functions $t_a$. 

\medskip

\fbox{
\begin{minipage}{0.9\textwidth}
\textbf{Arc flowing function definition}

\ligne 

Given an arc flow $\boldsymbol{Y}_a$ in $\mathcal M(R\times \mathbb{R})$, the flow of users following route $r$ and leaving the arc $a$ on the measurable time subset $J$ is:
\begin{equation}
\psi_a^r(\boldsymbol{Y}_a)(J):=\psi_a(\boldsymbol{Y}_a)(\{r\}\times J):=\left\{
\begin{array}{cl}
Y_a^r(H_a(Y_a)^{-1}(J)) & \mbox{if $a\in r$} \\ 0 & \mbox{if not,}\end{array}\right. \end{equation}
where $Y_a$ denotes the total flow on arc $a$,defined by $ \sum_{r':\,r'\ni a}Y_a^{r'}$.
\end{minipage}
}

\begin{claim}
By countable unions and intersections, $\psi_a$ definition can be extended on all measurable sets of $R\times\mathbb{R}$.
\end{claim}

The proof is straightforward. It uses the continuity of $H_a(Y_a)$ (Assumption \ref{ass:cont}). Hence, $\psi_a$ is well-defined.

The $\psi_a$ definition looks complicated, but actually, it is the natural definition of the flow leaving the arc $a$. Indeed, $H_a(Y_a)^{-1}(J)$ is the set of entrance times that allows to leave the arc $a$ during the set of exit times $J$ (according to the arc travel time function $t_a$) when a flow $Y_a$ goes through $a$. Hence, the volume of users leaving $a$  during $J$ and following route $r$ is nothing else than the volume of the flow that has entered $a$ during $H_a(Y_a)^{-1}(J)$. 

Now, let us state a technical lemma, used in the proof of Lemma \ref{lem:routearc}.

\begin{lemma}\label{lem:tech}
For all $h\in\mathbb{R}$ and $\boldsymbol{Y}_a\in\mathcal{M}(\mathbb{R})$, we have 
$$\psi_a(\boldsymbol{Y}_a)|_{H_a(Y_a)(h)}=\psi_a(\boldsymbol{Y}_a|_h)|_{H_a(Y_a)(h)}$$
and
\begin{equation}\label{eq:time}\psi_a(\boldsymbol{Y}_a)|_{h+t_{\min}}=\psi_a(\boldsymbol{Y}_a|_h)|_{h+t_{\min}}.\end{equation}
\end{lemma}

\begin{proof}
As soon as the first equality is true, the second one is also true, as a consequence of Claim \ref{cl:rest} and of Assumption \ref{ass:noinfspeed}.

Let us prove the first equality. Fix $h\in\mathbb{R}$, $Y,Y'\in\mathcal{M}(\mathbb{R})$ and $E$ a measurable subset of $\mathbb{R}$. We first prove two properties.

\medskip

\noindent {\em Property 1}: $H_a(Y|_h)^{-1}(E)\,\cap\, ]-\infty,h]=H_a(Y)^{-1}(E)\,\cap\,]-\infty,h]$.

\medskip

Indeed, for $h'\leq h$, we have 
$H_a(Y|_h)(h')=H_a(Y|_h|_{h'})(h')=H_a(Y|_{h'})(h')=H_a(Y)(h')$ with the help of Claim \ref{cl:rest} for the second equality and of Assumption \ref{ass:causality} for the first and third equalities. 

\medskip

\noindent {\em Property 2}: If $E\subseteq]-\infty, H_a(Y)(h)]$, and if $Y'\leq Y$, then $Y'\left(H_a(Y)^{-1}(E)\,\cap\, ]h,+\infty[ \right)=0$.

\medskip

Indeed, let $h'\in H_a(Y)^{-1}(E)\,\cap\, ]h,+\infty[$. We have $h'>h$ and $H_a(Y)(h')\leq H_a(Y)(h)$. According to Assumption \ref{ass:strictfifo}, we have then $Y[h,h']=0$, and hence $Y'[h,h']=0$.

\medskip

Take now $h\in\mathbb{R}$, $\boldsymbol{Y}\in\mathcal{M}(R\times\mathbb{R})$, $r\in R$ and $J$ a measurable subset of $\mathbb{R}$. Define $E:=J\cap]-\infty,H_a(Y_a)(h)[$. The set $E$ is a measurable subset of $\mathbb{R}$ and it is such that $E\subseteq]-\infty, H_a(Y_a)(h)]$. Note that $Y^r\leq Y_a$ when $a\in r$.

$$\begin{array}{rcll} 
\psi_a(\boldsymbol{Y}_a)|_{H_a(Y_a)(h)}(\{r\}\times J) & =  & Y^r_a(H_a(Y_a)^{-1}(E)) & \mbox{(by definition)}\\ &=&Y^r_a(H_a(Y_a)^{-1}(E)\,\cap\,]-\infty,h]) & \\
& &\quad+Y_a^r(H_a(Y_a)^{-1}(E)\,\cap\,]h,+\infty[) & \mbox{(since $Y^r$ is a measure)}\\
 &= & Y^r_a(H_a(Y_a)^{-1}(E)\,\cap\,]-\infty,h]) & \mbox{(according to Property 2)}\\
 &= & Y^r_a(H_a(Y_a|_h)^{-1}(E)\,\cap\,]-\infty,h]) & \mbox{(according to Property 1)}\\
 &= & Y^r_a|_h(H_a(Y_a|_h)^{-1}(E)\,\cap\,]-\infty,h]) & \mbox{(by definition of $|_h$)}\\
 &= & Y_a^r|_h(H_a(Y_a|_h)^{-1}(E)\,\cap\,]-\infty,h]) & \\
 & & \quad+Y_a^r|_h(H_a(Y_a|_h)^{-1}(E)\,\cap\,]h,+\infty[) & \mbox{(by definition of $|_h$)}\\
 &= & Y^r_a|_h(H_a(Y_a|_h)^{-1}(E)) & \mbox{(since $Y^r|_h$ is a measure)}\\
 &= & \psi_a(\boldsymbol{Y}_a|_h)|_{H_a(Y_a|_h)(h)}(\{r\}\times J) & \mbox{(by definition).}
\end{array}$$
\end{proof}

\subsection{Flowing model}\label{subsec:flowm}

A flow on an arc, as we have explained in the previous subsection, is a vector of measures in $\mathbb{R}$. To explain how flows $\boldsymbol{X}$ on routes induce flows $\boldsymbol{Y}=\left(Y^r_a\right)_{ a\in A , r\in R}$ on arcs $a$ and along routes $r$, we need to precisely define the outflow of $\boldsymbol{X}$ over the network. 

\medskip

\fbox{
\begin{minipage}{0.9\textwidth}
\textbf{Outflow on a network}

\ligne
An {\em outflow} of $\boldsymbol{X}$ is a collection of flows $\boldsymbol{Y}=\left(Y^r_a\right)_{ a\in A , r\in R}$, seen as a measure on $A\times R \times \mathbb{R}$ such that for every $r=a_1a_2\ldots a_n$:

\begin{equation}\label{eq:outflow}\left\{
\begin{array}{rcl}
Y^r_{a_1} & = & X_r  \\ 
Y^r_{a_i} & = & \psi_{a_i}^r(\boldsymbol{Y}_{a_{i-1}})\quad\mbox{\textup{for $i=2,\ldots,n$}} \\
Y^r_{a} & = & 0 \quad\mbox{\textup{if $a \notin r$}}\\
\end{array}\right.\end{equation}
\end{minipage}
}

\medskip

The previous definition is to be interpreted as such. Along a route $r$ the flow of users entering an arc $a$ and following a route $r$ corresponds to the flow entering the route if $a$ is the first arc of $r$, and otherwise to the flow leaving the previous arc. In the second case, the flow leaving the previous arc can be related to the one entering by an arc flowing function. Naturally, a flow on a route $r$ induces no flow on an arc $a$ if $a\notin r$. 

Now, we are in position to explain how it is possible to derive flows on arcs from flows on routes. 

\begin{lemma}\label{lem:routearc}
Fix $k\in\mathbb{N}$.
Given a measure $\boldsymbol{X}\in\mathcal{M}(R\times I)$ (the flows on the routes), there exists a unique $\boldsymbol{Y}=\left(Y^r_a\right)_{a\in A, r\in R}$ in $\mathcal{M}(A\times R\times\mathbb{R})$ such that for all route $r=a_1a_2\ldots a_n$

$$(E_k)\quad\left\{
\begin{array}{rcl}
Y^r_{a_1} & = & X_r|_{kt_{\min}}  \\ 
Y^r_{a_i} & = & \left. \psi_{a_i}^r(\boldsymbol{Y}_{a_{i-1}})\right|_{kt_{\min}}\quad\mbox{\textup{for $i=2,\ldots,n$}} \\
Y^r_{a} & = & 0 \quad\mbox{\textup{if $a \notin r$}}\\
\end{array}\right.$$

Moreover, for any $a$, the map $\displaystyle \phi^k_a:\boldsymbol{X} \mapsto Y_a:=\sum_{r:\,a\in r} Y^r_a$, where $(\boldsymbol{Y}_a)_{a\in A}$ is the solution of $(E_k)$, is continuous.
\end{lemma}

Lemma \ref{lem:routearc} informally says that it is possible to construct a sequence of measures on $A\times R \times \mathbb{R}$, with each of its element representing the progressive propagation of the flow of users over the network, with a time step of $t_{min}$. The proof of the lemma relies on Assumption \ref{ass:noinfspeed} (no infinite speed on an arc), which highlights the crucial importance of this assumption in our approach.

\begin{proof}
The proof works by induction on $k$. For $k=0$, define $Y^r_a:=0$ for all $r$ and $a$. And there is no other solution.

Suppose now that $k\geq 0$ and that we have proved the lemma till $k$. 

\medskip

\noindent{\em Existence and continuity}:  Let $\boldsymbol{Y}'=\left(Y'^{r}_a \right)_{a\in A,r\in R}$ be the solution of $(E_k)$. We want to prove that $(E_{k+1})$ has a solution. Define $\boldsymbol{Y}$ for all routes $r=a_1a_2\ldots a_n$ by
$$\begin{array}{rcll}

Y^r_{a_1} & := & X_r|_{kt_{\min}} &  \\ 
Y^r_{a_i} & := &\left. \psi_{a_i}^r(\boldsymbol{Y}'_{a_{i-1}})\right|_{kt_{\min}} & \mbox{for $i=2,\ldots,n$} \\
Y^r_a & := & 0 & \mbox{if $a\notin r$}
\end{array}$$

According to this definition, $\boldsymbol{Y}$ depends continuously on $\boldsymbol{X}$.

Note that, according to Claim \ref{cl:rest}, we have then for all $a\in A$, $r\in R$
\begin{equation}\label{eq1}
\boldsymbol{Y}'^r_a=\boldsymbol{Y}_a^r|_{kt_{\min}}
\end{equation}

We check that $\boldsymbol{Y}$ is solution of $(E_{k+1})$. The first and the last equalities of $(E_{k+1})$ are straightforward. Let us check the second one. Let $r$ be in R and $a'a\subset r$.

$$\begin{array}{rcll}
\boldsymbol{Y}_a^r & = & \psi_{a'}^r\left.\left(\boldsymbol{Y}'_{a'}\right)\right|_{(k+1)t_{\min}} & \mbox{(by definition of $\boldsymbol{Y}$)}\\
& = & \psi_{a'}^r\left.\left(\left.\boldsymbol{Y}_{a'}\right|_{kt_{\min}}\right)\right|_{(k+1)t_{\min}} & \mbox{(according to Equation (\ref{eq1}))}\\
& = & \psi_{a'}^r\left.\left(\left.\boldsymbol{Y}_{a'}\right|_{(k+1)t_{\min}}\right)\right|_{(k+1)t_{\min}} & \mbox{(according to Equation (\ref{eq:time}) of Lemma \ref{lem:tech})}\\
& = & \psi_{a'}^r\left.\left(\boldsymbol{Y}_{a'}\right)\right|_{(k+1)t_{\min}} & \mbox{(according to Assumption \ref{ass:causality}).}
\end{array}$$

\medskip

\noindent{\em Uniqueness}:  Assume that we have two collections $\boldsymbol{Y}$ and $\boldsymbol{Z}$ solution of $(E_{k+1})$. Yet, $\boldsymbol{Y}|_{kt_{\min}}$ and $\boldsymbol{Z}|_{kt_{\min}}$ are solutions of $(E_k)$. Hence, by induction, 
\begin{equation}\label{YZ}
\left(\boldsymbol{Y}_a|_{kt_{\min}}\right)_{a\in A}=\left(\left.\boldsymbol{Z}_a\right|_{kt_{\min}}\right)_{a\in A}
\end{equation} 
We can write the chain of equalities for any $a\in A$
$$\begin{array}{rcll}
\boldsymbol{Y}_a^r & = & \psi_{a'}^r\left.\left(\boldsymbol{Y}_{a'}^r\right)\right|_{(k+1)t_{\min}} & \mbox{(since $\boldsymbol{Y}$ is solution of $(E_{k+1}))$}\\

 & = & \left.\psi_{a'}^r\left(\left.\boldsymbol{Y}_{a'}\right|_{kt_{\min}}\right)\right|_{(k+1)t_{\min}}& 
 \mbox{(according to Equation (\ref{eq:time}) of Lemma \ref{lem:tech})}\\
 & = & \left.\psi_{a'}^r\left(\left.\boldsymbol{Z}_{a'}\right|_{kt_{\min}}\right)\right|_{(k+1)t_{\min}} & \mbox{(according to Equation (\ref{YZ}))}\\
 & = & \boldsymbol{Z}_a^r & \mbox{(since $\boldsymbol{Z}$ is solution of $(E_{k+1})$).}
 \end{array}$$
\end{proof}

We are now in position to state and prove the main result of the subsection.

\begin{proposition}\label{prop:routearc}
Given the flow $\boldsymbol{X}\in\mathcal{M}(R\times I)$ of the users on the routes, there is a unique outflow $\boldsymbol{Y}\in\mathcal{M}(A\times R\times\mathbb{R})$ of users on the arcs satisfying system (\ref{eq:outflow}). For each arc $a$, the outflow $\boldsymbol{X}\mapsto\boldsymbol{Y}_a$ is given by a continuous map.
\end{proposition}

\begin{proof}
Recall that $N=\sum_{r\in R}X_r(I)$ is the total number of users.
Let $\tau:=\max_{x\in[0,N]}t_{\max}(x)$. According to Assumption \ref{ass:end}, for any route $r=a_1a_2\ldots, a_n$, a direct induction on $i$ leads to $\boldsymbol{Y}_{a_i}=\left.\boldsymbol{Y}_{a_i}\right|_{i\tau}$ (nobody leaves arc $a_i$ after $i\tau$). Hence, any outflow $\boldsymbol{Y}$ solution of (\ref{eq:outflow}) is solution of Equation ($E_k$) for a $k$ big enough. Existence, continuity, and uniqueness are consequence of Lemma \ref{lem:routearc}. 
\end{proof}

According to Proposition \ref{prop:routearc}, we can define for each arc $a\in A$ a continuous map $\phi_a:\mathcal{M}(R\times I)\rightarrow\mathcal{M}(\mathbb{R})$ such that given a flow $\boldsymbol{X}\in\mathcal{M}(R\times I)$ of the users on the routes, $\phi_a(\boldsymbol{X})$ is the total flow $Y_a\in\mathcal{M}(\mathbb{R})$ of users on the arc $a$.

\section{Existence of an equilibrium}

\label{sec:cont}

\begin{proof}[Proof of Theorem \ref{thm}] Theorem \ref{thm} is a consequence of the following lemma (Lemma \ref{lem:cont}) and of Theorem \ref{thm:kh}.
\end{proof}

We have defined $t_r$ in terms of $t_a$ in the Introduction by Equation (\ref{eq:tr}). It is straightforward to check that we have also for a route $r=a_1,\ldots,a_n$
\begin{equation}\label{tr}
t_r(\boldsymbol{X})(h)=\left(H_{a_n}\left(\phi_{a_n}(\boldsymbol{X})\right)\circ\ldots\circ H_{a_1}\left(\phi_{a_1}(\boldsymbol{X})\right)\right)(h)-h\quad\mbox{for all }h\in I.
\end{equation}

\begin{lemma} \label{lem:cont}  If the arc travel time functions $t_a:\mathcal{M}(\mathbb{R})\rightarrow\mathcal{C}(\mathbb{R})$ satisfy the five assumptions of Subsection \ref{subsec:time} for each arc $a\in A$, and if the arc flows satisfy the three assumptions of the flowing model (Subsection \ref{subsec:flowm}), then the route travel time functions $t_r:\mathcal{M}(I)\rightarrow\mathcal{C}(\mathbb{R})$ are continuous for each route $r\in R$.
\end{lemma}
 
 
\begin{proof}
Since from Proposition \ref{prop:routearc}, $\phi_a:\mathcal{M}(R\times I)\rightarrow\mathcal{M}(R\times\mathbb{R})$ and $H_a:\mathcal{M}(R\times\mathbb{R})\rightarrow\mathcal{M}(\mathbb{R})$ are continuous, it remains to prove that
the compositions in Equation (\ref{tr}) keep continuity. It is a consequence of the following property.

\medskip

{\em Let $I'$ be a closed interval of $\mathbb{R}$ and $f:\mathcal{M}(\mathbb{R})\rightarrow\mathcal{C}(\mathbb{R})$ and $g:\mathcal{M}(I')\rightarrow\mathcal{C}(I')$ be two continuous functions. Assume moreover that $f(Y)$ is uniformly continuous for all $Y\in\mathcal{M}([0,\tilde{H}])$. Then $Y\mapsto f(Y)\circ g(Y)$ is continuous.}

\medskip

Indeed, let $\epsilon>0$ and $Y\in\mathcal{M}([0,\tilde{H}])$. 

According to the continuity of $f$, there is an $\eta_1>0$ such that $\rho(Y,Y')\leq\eta_1$ implies $||f(Y)-f(Y')||_{\infty}\leq\epsilon/2$.

According to the uniform continuity of $f(Y)$ on the image of $g(Y)$, which is compact, there is an $\eta_2>0$ such that for all $h,h'\in[0,\tilde{H}]$, when $|h-h'|\leq\eta_2$, we have $|f(Y)(h)-f(Y)(h')|\leq\epsilon/2$.

According to the continuity of $g$, there is an $\eta_3>0$ such that $\rho(Y,Y')\leq\eta_3$ implies $||g(Y)-g(Y')||_{\infty}\leq\eta_2$.

Now define $\eta:=\min(\eta_1,\eta_3)$. For all $Y'\in\mathcal{M}([0,\tilde{H}])$ such that $\rho(Y,Y')\leq\eta$, we have 
$$||f(Y)\circ g(Y)-f(Y')\circ g(Y')||_{\infty}
\leq ||f(Y)\circ g(Y)-f(Y)\circ g(Y')||_{\infty}+||f(Y)\circ g(Y')-f(Y')\circ g(Y')||_{\infty}
\leq \epsilon/2+\epsilon/2\leq\epsilon.$$ Thus the property. Then, by simple induction, the continuity of $\phi$ is straightforward.
\end{proof}

\section{Applications to the dynamic Wardrop assignment} \label{sec:applications}
Our result is fairly general and notably apply to most of the problems of dynamic assignment at equilibrium found in the transportation literature. Those models, although commonly used in practice for transportation planning, lack of theoretical foundations and results of existence have been established only in very restrictive cases.

In this section, we show that the existence of a solution to the most common dynamic assignment problem, the so-called dynamic Wardrop assignment problem, has a solution under the general travel time assumptions we stated earlier. Then, two common travel time models in the transportation literature are reviewed and and it is shown that they are natural travel times functions in the sense stated above. 



\subsection{Dynamic Wardrop assignment with predetermined departure times}
The simplest assignment model can be formulated as such. Consider a travel demand, described by flows between each origin destination pair, and assume each of them is allowed to choose its travel route, but not its departure time. We study the possible assignments of the traffic flows on the routes connecting each of the $OD$ pair. The question is the following: is there an assignment such that no route is assigned at a time $h$ with a non zero flow of vehicles if there are routes with smaller travel times ? Such an assignment is said to verify the \emph{Dynamic Wardrop Principle}. Note that the terminology in the transportation literature is variable from one authors to another and that what we call dynamic Wardrop assignment, is also termed as user equilibrium assignment (\cite{FBS93} or \cite{ZHM00}).

A formal statement of the dynamic assignment problem is presented below. Before doing so let us raise a few comments on the mathematical nature of traffic flows in transportation model compared to ours. Existing models represent flows by measurable functions, whereas our formulation is based on measures on $I$, so it is useful to identify each element of $L(I,\mathbb{R}_{+})$, the set of positive measurable functions on $I$, with an element of $\mathcal{M}(I)$. Thus to a flow $x \in L(I,\mathbb{R}_{+})$, we associate the measure $X$ defined by $X([0,h])=\int^h_0 x(h') dh$.



The dynamic Wardrop assignment problem can now be formulated. Consider a directed graph $G=(V,A)$, with arc travel time functions $(t_a)_{a\in A}$ and  an {\em OD matrix} $(q_{od})_{o\in V,d \in V}$, each element of the matrix being a function in $ L(I,\mathbb{R_+})$. We define the route travel time functions $(t_r)$ as in the previous sections (with the same flowing model). An \emph{assignment of the traffic} is an element $\boldsymbol{x}=(x_r)$ of $L(I,\mathbb{R_+})^R$ such that $\sum_{r\in R_{o,d}}x_r(h)=q_{od}(h)$ for all $(o,d)\in V\times V$ and $h\in I$, with $R_{o,d}$ denoting the set of routes connecting $o$ to $d$. 

\medskip

\fbox{
\begin{minipage}{0.9\textwidth}
\textbf{Dynamic Wardrop Assignment Problem} 

\ligne
Find an assignment $x\in L(I,\mathbb{R_+})^R$ such that whenever $r,r'\in R_{o,d}$
$$x_r(h)>0\Rightarrow t_r(\boldsymbol{X})(h)\leq t_{r'}(\boldsymbol{X})(h),\mbox{ for a.e. $h\in I$}$$
\end{minipage}}

\medskip

\begin{theorem}\label{th:ass}
Given a directed graph $G=(V,A)$ with arc travel time functions $(t_a)_{a\in A}$ satisfying Assumptions \ref{ass:cont}-\ref{ass:causality} and given an $OD$ matrix, there is a Wardrop assignment.
\end{theorem}

The proof is an application of Theorem \ref{thm}. We consider a distribution of users $U$ on the set $\mathcal{RC}$ of continuous utility functions of the following type
\begin{equation}
\hat{u}_{h^{\ast},od}(\boldsymbol{t})(r,h) = \left\lbrace
\begin{array}{lc}
 -t_r(h)& \mbox{if $h=h^{\ast}$ and $r\in \text{od}$} \\
 -\infty& \mbox{otherwise,}
\end{array} \right.
\end{equation} 

The interpretation is straightforward : each user is characterized by a departure time $h^{\ast}$ he will always prefer, and an origin-destination pair $od$ on which he will always travel. The utility of a travel decision is limited to the travel time on the route. 

The set $\mathcal RC$ can be identified as $V\times V\times I$, introducing the continuous 
mapping\footnote{
Recall that when $E$, $F$ and $G$ are three topological spaces and $f:X\times Y \rightarrow G$ is a continuous map, then the map $F:X \mapsto C(Y,Z)\ :\ F(x)=f(x,y)$ is continuous. Apply this proposition to $\left((od,h^{*}),\boldsymbol{t} \right) \mapsto \hat{u}_{h^{\ast},od}(\boldsymbol{t})$, which is trivially continuous with the chosen topologies. Then $(od,h^{*})\mapsto \hat{u}_{h^{\ast},od}$ is continuous.}  $(od,h^{*})\mapsto \hat{u}_{h^{\ast},od}$. Consequently $\mathcal RC$ is compact as the image of a compact by a continuous function, and hence (Borel) measurable.
According to the context a measure on $\mathcal RC$ is seen either as a measure on $\mathcal{C}(\mathcal{S}_{R\times I})$, or as a collection of measures $(U_{od})_{o\in V,d\in V}$ on $I$. The latter point of view is of particular interest because of the following proposition:

\begin{proposition}
\label{pr:flowrod}
When $U$ is a measure on $\mathcal{RC}$ seen as measure on $\mathcal{C}(\mathcal{S}_{R\times I})$, the equilibrium flow $\boldsymbol{X}$ verifies:
\begin{equation}
U_{od} = \sum_{r\in R_{o,d}} X_r
\end{equation}
\end{proposition}

\begin{proof}
Consider a measure $U$ on $\mathcal{RC}$ $U$ on $\mathcal{C}(\mathcal{S}_{R\times I})$ such that $U(\mathcal{C}(\mathcal{S}_{R\times I}))=U\{\hat{u}_{h,od}:\,h\in I, od\in V\times V\}$. Let $D$ be an associated Nash equilibrium. Recall that $\boldsymbol{X}:=D_{R\times I}$ and $U=D_{\mathcal{C}(\mathcal{S}_{R\times I})}$. Then for all measurable subsets $E$ of $I$:
$$\begin{array}{rcll}
\boldsymbol{X}(R_{o,d}\times E)&=&D_{R\times I}(R_{o,d}\times E) &\mbox{(by definition of $\boldsymbol{X}$ )}\\
&=&D(R_{o,d}\times E \times \mathcal{C}(\mathcal{S}_{R\times I})) &\mbox{(by definition of a margin)}\\
&=&D(R_{o,d}\times E \times \mathcal{RC} )&\mbox{($U$ is a measure on $\mathcal{RC}$) }\\
&=&D(R_{o,d}\times E \times \{ \hat{u}_{h^{\ast},od}	\mbox{ such that } h^{\ast}\in E \})&\mbox{(D is an equilibrium measure) }\\
&=&D(R\times I \times \{ \hat{u}_{h^{\ast},od}	\mbox{ such that } h^{\ast}\in E \})&\mbox{(idem) }\\
&=&U(\{ \hat{u}_{h^{\ast},od}	\mbox{ such that } h^{\ast}\in E \})&\mbox{(idem) }\\
&=&U_{od}(E)&\mbox{(identifying $\mathcal{RC}$ with $V\times V \times I$) }\\
\end{array}$$

\end{proof}

Note that in the dynamic Wardrop assignment problem as formulated above, one is only interested in measures $U$ such that there exists a collection $(q_{od})_{o\in V,d\in V}$ of positive measurable functions defining $U$ through equation
$U_{od}(E)=\int_E q_{od}(h)dh$, for all measurable subsets $E$ of $I$. This is exactly the set of absolutely continuous measures \footnote{
$\nu$ is said to be absolutely continuous with respect to $\mu$ if $\nu(A) = 0$ for every set $A$ for which $\nu(A) = 0$. In finite dimensional spaces, the absolutely continuous measures with respect to the Lebesgue measure are exactly the ones that have a density.
}
 on $V\times V\times I$ with respect to the Lebesgue measure. 

Theorem \ref{thm} tells us that there is an equilibrium, but this equilibrium is a measure $D$ leading to flows $X_r:=D_{R\times I}(\{r\}\times I)$ that might not have a derivative that is an element of $L(I,\mathbb{R_+})$. Our equilibrium might not be an equilibrium is the sense above.

Fortunately, we have the following lemma.

\begin{lemma}\label{lem:nodirac}
Let $U$ be a measure on $\mathcal{RC}$, seen as measure on $\mathcal{C}(\mathcal{S}_{R\times I})$. When $U$ is absolutely continuous, every equilibrium flows $\boldsymbol{X}$ is also absolutely continuous.
\end{lemma}
\begin{proof}
Consider an absolutely continuous measure $U$ on $\mathcal{C}(\mathcal{S}_{R\times I})$ such that $U(\mathcal{C}(\mathcal{S}_{R\times I}))=U\{\hat{u}_{h,od}:\,h\in I, od\in V\times V\}$ and let $\boldsymbol{X}$ be an associated Nash equilibrium. According to Proposition \ref{pr:flowrod}:
$$U_{od} = \sum_{r\in R_{o,d}} X_r$$
Then if we have $E$ a measurable subset of $I$ such that $U_{od}(E)=0$, for all $r\in R_{o,d}$ we have $X_r(E)=0$. Thus, absolute continuity of $U$ implies absolute continuity of $\boldsymbol X$.
\end{proof}

\begin{proof}[Proof of Theorem \ref{th:ass}]
Assume we are given an OD matrix $(q_{od})$. Consider $U$ a measure on $\mathcal{C}(\mathcal{S}_{R\times I})$ such that
\begin{itemize}
\item $U(\mathcal{C}(\mathcal{S}_{R\times I}))=U\{\hat{u}_{h,od}:\,h\in I, od\in V\times V\}$ and
\item for a given pair $od\in V\times V$ and any measurable subset $J\subseteq I$, we have $U\{\hat{u}_{h,od}:\,h\in J\}=\int_{h\in J}q_{od}(h)dh$.
\end{itemize}
We have just encoded our OD matrix as a measure on the set of users. Note that $U$ is absolutely continuous.

According to Theorem \ref{thm}, there exists a Nash equilibrium $D$, and according to Lemma \ref{lem:nodirac} the equilibrium flows
 $\boldsymbol{X}:=D_{R \times I}$ is absolutely continuous regarding the Lebesgue measure. Hence $\boldsymbol{X}$ admits a Radon Nikodym derivative, which we will denote $\boldsymbol{x}$. Let $h\in I$ and take any route $r$ such that $x_r(h)>0$. Let $od$ the origin-destination pair connected by $r$. The proof proceeds in two steps. First, we show that whenever $x_r$ is continuous in $h$, $x_r(h)>0\Rightarrow t_r(\boldsymbol{X})(h)\leq t_{r'}(\boldsymbol{X})(h)$ for all $r'\in R_{o,d}$. Then, we show that this inequality holds almost everywhere.
\medskip

\textbf{First step.} Let $h\in I$ be such that $\boldsymbol{x}$ is continuous in $h$. Now, take any route $r$ such that $x_r(h)>0$. Let $od$ the origin-destination pair connected by $r$. For all $\epsilon>0$, we have $X_r([h-\epsilon ; h+\epsilon])>0$, which can be rewritten $D\left(\{r\}\times\{h'\}\times\hat{u}_{h',od}:\,h'\in[h-\epsilon,h+\epsilon]\right)>0$. 
Therefore, we know that for all $\epsilon>0$, there is $h'\in[h-\epsilon,h+\epsilon]$ such that $\hat{u}_{h',od}(\boldsymbol{t}(\boldsymbol{X}))(r',h'')\leq\hat{u}_{h',od}(\boldsymbol{t}(\boldsymbol{X}))(r,h')$ for all $h''\in I$ and $r'\in R$, or, directly in terms of route travel times: 
$$\mbox{for all $\epsilon>0$, there is $h'\in[h-\epsilon,h+\epsilon]$ such that } t_{r'}(\boldsymbol{X}))(h')\geq t_r(\boldsymbol{X})(h')\mbox{ for all $r'\in R_{o,d}$}.$$
By continuity of $h\mapsto t_r(\boldsymbol{X})(h)$, we get the required inequality.
\medskip

\textbf{Second step.} For a given $r$ consider $E$ the set of point such that $x_r(h)>0$ and $t_r(X)(h)>t_{r'}(X)(h)$ for a $r'$ on the same $OD$ pair as $r$. From the previous paragraph $x_r$ is discontinuous in every $h\in E$. $E$ is measurable as $t_r$, $t_{r'}$ and $x_r$ also are. Now assume $\mu(E)=\epsilon\neq 0$, denoting $\mu$ the Lebesgue measure on $\mathbb{R}$. Then, $x_r$ being measurable, there exists a set $K$ such that the measure of its complementary $\mu(K^{c})<\epsilon/2$ and $x_r$ is continuous in every $h\in K$ (Lusin Theorem \cite{LUS12}). So $K \cap E\neq \emptyset$, a contradiction. Hence $\mu(E)=0$.

\medskip

Thus, the required inequality is valid almost everywhere. 

\end{proof}

\subsection{Two arc travel times models}
The present subsection exposes two arc travel times models under which the existence of a Wardrop assignment has been shown, respectively by Zhu and Marcotte \cite{ZHM00} (although with slightly more specific assumptions) and Mounce \cite{MOU06} (although only on specific networks). We show here that both models verifies the Assumptions (\ref{ass:cont}-\ref{ass:causality}), and thus that those two existence results are direct consequences of Theorem \ref{thm}.
\subsubsection{Arc performance travel times}
In an arc performance model (see for instance Friez et al \cite{FBS93}), travel time on an arc is assumed to depend on the volume of traffic on that same arc. More precisely a \emph{delay function} $D_a:\mathbb{R}_+:\rightarrow \mathbb{R}_+^{\ast}$ taking in input a volume of traffic and returning a travel time such that $D_a$ is associated to each arc. It is assumed to be continuous, strictly increasing, positive and leading to a fifo property. Formally, the travel time model is defined by the following system:
\begin{equation}
\label{eq:ttzhm}
\left\lbrace
\begin{array}{lcl}
 \displaystyle V_a(Y_a)(h)&=&Y_a([-\infty,h])-Y_a\left(H_a(Y_a)^{-1}\left([-\infty,h]\right)\right)\\
 \displaystyle H_a(Y_a)(h)&=&h+D_a\left(V_a(Y_a)(h)\right)\\
\end{array} \right.
\end{equation}
We will refer to such a travel time function hence defined as \emph{arc performance travel time functions}, in line with the classical transportation terminology. Remark that only Assumptions \ref{ass:cont} and \ref{ass:causality} (continuity and causality) are not straightforward. But it is possible to prove them.

\begin{proposition}\label{pr:ARTT}
Assuming that all users enter the network during the time interval $I$, the functions $H_a$ are well-defined by Equations \ref{eq:ttzhm} and satisfy Assumption  \ref{ass:cont}-\ref{ass:causality}.
\end{proposition}
\begin{proof}
The proof works very similarly to the one of Lemma \ref{lem:routearc}. 

Define $d_{\min}:=D_a(0)$. Then necessarily, for $h\in]-\infty,d_{\min}]$, we have $V_a(Y_a,h)=Y_a]-\infty,h]$. Hence, we have defined $V_a(Y_a,h)$ and $H_a(Y_a)(h)$ for all $h\in]-\infty,d_{\min}]$.

Since we have $H_a(Y_a)^{-1}\left(]-\infty,2d_{\min}]\right)\subseteq]-\infty, d_{\min}]$, we can now define $V_a(Y_a,h)$ and $H_a(Y_a)(h)$ for all $h\in]-\infty,2d_{\min}]$. And so on: for all $k\in\mathbb{N}$, we have $H_a(Y_a)^{-1}\left(]-\infty,kd_{\min}]\right)\subseteq]-\infty, (k-1)d_{\min}]$, we can define $V_a(Y_a,h)$ and $H_a(Y_a)(h)$ for all $h\in]-\infty,(k-1)d_{\min}]$.

Since we work by necessity and sufficient conditions, we have existence and uniqueness. Moreover, the map $Y_a\mapsto V_a(Y_a,\cdot)$ is clearly continuous.

It remains to prove causality, that is, $H_a(Y_a|_h)(h)=H_a(Y_a)(h)$. Without loss of generality, we can assume that there is a integer $k_h$ such that $h=k_hd_{\min}$. Actually, a direct induction on the index $k$ above, for $k=1$ to $k=k_h$, proves that whenever $h'\leq h$, we have $H_a(Y_a|_h)(h')=H_a(Y_a)(h')$. 
\end{proof} 


\subsubsection{Bottleneck travel times}
Mounce uses the punctual bottleneck model to represent travel time on an arc. The arc travel time for an entrance time $h$ is the sum of a constant travel time and a bottleneck delay. The delay arises from a limit $K$ on the arc outcoming flow, refer to as the capacity of the arc. When traffic flow exceeds this capacity, a punctual queue starts to form at the exit of the arc. In \cite{MOU06}, Mounce exposes how to express the travel time on an arc as a function of the cumulated volume at entrance and shows it is continuous and respect the fifo condition. 

\subsection{Further results}
It's quite clear that Theorem \ref{thm} can be applied to far more general models than the dynamic Wardrop assignment. Allowing users to choose their departure times as well as their routes, in a similar manner as done by Arnott et al or Friez et al \cite{APL89,FBS93}, is the obvious and natural next step. But much more complex models actually fit in our framework. 

On the demand side, the set of utility functions $\mathcal{C}(\mathcal{S}_{R\times I})$ allows an incredibly large set of variations. For instance, utilities that varies non linearly with travel time can be considered. This is of particular importance, as studies show it is empirically relevant (see for instance \cite{KOP81}). Second, road pricing strategies can be embedded in the utility functions, by adding maluses on specific routes. 

On the supply side, the assumptions we considered are very weak and also include a wide range of particular models. The two specific cases we consider are very simple, but it is likely that more complex traffic models, such as the ones inspired from fluid mechanics, would also fit in our framework.

\bibliographystyle{plain} 
\bibliography{biblio} 
\end{document}